\documentclass[12pt]{article}
\usepackage{orcidlink}
\usepackage{graphicx} 
\usepackage{graphics}
\usepackage{fancybox}
\usepackage{amsmath,latexsym,psfrag,epsf,epsfig,amssymb,rotating}
\usepackage{color}
\usepackage{multirow}
\usepackage{url}
\usepackage{natbib}
\usepackage{geometry}
\usepackage{fullpage}
\usepackage{subfigure}
\usepackage{bm}
\usepackage{authblk}
\usepackage[english]{babel}
\usepackage[utf8x]{inputenc}
\usepackage{dsfont}
\usepackage{algorithm}
\usepackage{algorithmicx}
\usepackage{algpseudocode}
\usepackage{multirow}
\usepackage{authblk}

\providecommand{\newblock}{}
\usepackage{setspace}
\usepackage{url}

\newcommand{\pref}[1]{%
    \eqref{#1} \ifnum\count0=\pageref{#1}\relax%
    \else (page \pageref{#1})\fi}

\newcommand{\eref}[1]{%
        \eqref{#1}\ifnum\count0=\pageref{#1}\relax%
        \else {, p.\pageref{#1}}\fi}

\newcommand{\comment}[1]{}

\newcommand{\bfLambda}{\boldsymbol{\Lambda}}

\newcommand{\bfvartheta}{\boldsymbol{\vartheta}}
\newcommand{\bfvarphi}{\boldsymbol{\varphi}}

\newcommand{\bfSigma}{\boldsymbol{\Sigma}}

\newcommand{\bfmu}{\boldsymbol{\mu}}

\def\bfs{\mathbf{s}}

\def\bfx{\mathbf{x}}

\def\bfy{\mathbf{y}}
\def\bfY{\mathbf{Y}}
\def\bfz{\mathbf{z}}

\providecommand{\keywords}[1]
{
  \small	
  \textbf{\textit{Keywords---}} #1
}

\title{Multi-layer dissolution exponential-family models\\ for signed weighted networks}

\author[1]{Alberto Caimo\orcidlink{0000-0001-8956-7166}\thanks{Corresponding author.}}
\author[2,3]{Paolo Dalena \orcidlink{0000-0003-2920-9572}}
\author[1,4]{Isabella Gollini \orcidlink{0000-0002-7738-5688}}

\affil[1]{\small School of Mathematics and Statistics, University College Dublin, Ireland}
\affil[2]{\small University of Trieste, Italy}
\affil[3]{\small Institute for Maternal and Child Health IRCCS Burlo Garofolo, Trieste, Italy}
\affil[4]{\small Rinn AI, University College Dublin, Ireland}

\begin{document}
\maketitle

\begin{abstract}
Understanding the structure of signed weighted networks is essential for analysing complex systems in which relationships between nodes vary both in sign and strength. Despite significant advances in statistical network analysis, there is still a lack of models that can account for both the sign and the strength of interactions in networks. We introduce a multi-layer dissolution exponential random graph modelling framework that enables systematic assessment of structural balance effects. We adopt a fully-probabilistic hierarchical approach that partially pools information across layers, with parameters estimated via an adaptive approximate exchange algorithm. A simulation study shows that the algorithm recovers the layer-specific parameters with small bias. 
We demonstrate the flexibility and explanatory power of the proposed methodology through an application to the gene interaction network associated with dilated cardiomyopathy, in which a multi-layer signed structure is constructed by progressively thresholding interaction confidence scores, with edge signs derived from pairwise co-expression. The analysis shows that positive co-expression edges embedded in positive triangles are more likely to persist at higher confidence levels, while negative edges whose endpoints share positive partners, which form unbalanced triangles, are preferentially dissolved as the confidence threshold increases.

\end{abstract}


\section{Introduction} 

Networks provide a powerful framework for representing complex relational data, capturing not only the presence or absence of connections between entities, but also the strength and structure of those interactions. As a result, they offer valuable insights across a wide range of domains including genomics \citep{mas:et09}.
Indeed, in many empirical settings, relationships are both weighted and signed, making it essential to capture the interaction between edge strength and sign in order to understand the mechanisms shaping relational structures.

Statistical modelling of networks has traditionally focused on binary graphs, with exponential random graph models (ERGMs; \citet{fra:str86,lus:kos:rob13}) emerging as a flexible and widely used class of models for representing dependence structures in relational data. 
ERGMs allow researchers to model the probability of observing a network as a function of local configurations, such as density, 
triadic closure, and nodal degrees, linking micro-level interaction processes to global network patterns \citep{sni:pat:rob:han06}. 

Extensions to weighted networks \citep{kri12, wil:etal17, kri:but17, cai:gol20} have enabled the joint analysis of edge strength and structural features, providing richer insight into network formation dynamics. A recent systematic review by \cite{fan:whi24} comparing various weighted ERGM frameworks for neuroimaging network data found that the multi-layer dissolution ERGM proposed by \cite{cai:gol20} exhibited the most robust performance, distinguished by its ability to preserve informative edge weights, ease of implementation, and consistently strong results in simulation studies.

Another important line of research that has developed around ERGMs concerns the extension to multi-layer networks consisting of a common set of nodes incident on multiple types of relationships. \citet{wan13} reviewed the models introduced by \citet{laz:pat99} and presented two-layer extensions of the non-degenerate specifications of \citet{sni:pat:rob:han06}. A decisive advance in this area has recently been achieved through the approach of \citet{kri:koe:mar20}, which enables the joint modelling of associations across multiple network layers while accounting for structural properties both within and between layers. 

Signed networks are often subject to distinct structural constraints, such as structural balance \citep{car:har56}. Recently, \citet{fri:meh:thu:kau22} extended ERGMs to model simple (i.e., unweighted) signed network generative processes by adopting a categorical reference measure and  \citet{cai:gol26} developed  separable temporal ERGMs conditional on the interaction structure.

Despite these advances, there remains a need for models that can simultaneously capture signed and weighted interactions, test hypotheses such as structural balance, and allow for efficient parameter estimation while borrowing strength across related processes. To address this gap, we propose a Bayesian hierarchical multi-layer dissolution ERGM framework for signed weighted networks based on the approach of \cite{cai:gol20} enabling enhanced inference and interpretability of structural relational effects. 

The paper is structured as follows. In Section~\ref{sec:ergms}, we provide a brief overview of exponential random graph models. Section~\ref{sec:signed_nets} introduces signed networks and outlines the modelling approach proposed by \cite{fri:meh:thu:kau22}. Section~\ref{sec:multi_layer} introduces the dissolution process framework for jointly modelling signed and weighted processes conditional on the interaction process. It also presents several model specifications within the multi-layer framework introduced in the previous section. In Section~\ref{sec:Bayes}, we complete the modelling framework by introducing a Bayesian hierarchical structure for the layer-specific ERGM parameters. To sample from the intractable posterior distribution we develop an adaptive approximate exchange algorithm \citep{Murray06, cai:fri11} (see \citet{par:har18} for a recent review on Bayesian computation for intractable models), which updates the layer-specific parameters either layer by layer or jointly, with proposal covariances learned during the burn-in; the algorithm and the related computational considerations are described in the same section. Section~\ref{sec:sim} reports a simulation study that assesses parameter recovery and the effect of the hierarchical prior. In Section~\ref{sec:app}, we illustrate the proposed model through an analysis of a signed weighted gene interaction network associated with dilated cardiomyopathy, in which edge weights are given by interaction confidence scores and edge signs by pairwise co-expression, and we assess the fitted model through prior and posterior predictive checks. Concluding remarks are provided in Section~\ref{conc}.

\section{Exponential random graph models}\label{sec:ergms}

Exponential random graph models (ERGMs, \citealp{lus:kos:rob13}) are a flexible class of discrete exponential-family models that represent the probability distribution of a random network $\bfY$ as
$$
p(\bfy \mid \bfvartheta) = \exp\{\bfvartheta^\top \bfs(\bfy) - \kappa(\bfvartheta)\}\;h(\bfy),
$$
where $\bfvartheta$ is the parameter vector associated with the vector of sufficient network statistics $\bfs(\bfy)$, $\kappa(\bfvartheta)$ is the log-normalising constant, and $h(\bfy)$ is a reference measure including all terms not dependent on the parameter $\bfvartheta$ and specifies how probabilities are computed over the sample space. Each component of $\bfs(\bfy)$ typically encodes a structural feature of the network (e.g., edge density, transitivity, or degree heterogeneity), and the corresponding parameter $\vartheta_r$ measures its contribution to the likelihood on the log-odds scale. 

Several important new ERGM-based approaches have recently been introduced in the 
literature (see \cite{cai:gol23} for a recent review). 
We briefly focus on two in particular: valued-edge and multi-layer extensions.

For valued networks, ERGMs can be extended by defining an appropriate reference 
measure $h(\bfy)$ that captures the baseline distribution of edge weights. Following 
\citet{kri12}, this is often achieved by assuming that edge variables follow a positive 
support distribution such as the Poisson, in which case 
$h(\bfy) = \prod_{i<j} (y_{ij}!)^{-1},$ where $y_{ij}$ is the observed value
for the dyad $(i, j).$

Multi-layer ERGMs \citep{kri:koe:mar20} extend the exponential-family framework to account for network data observed across multiple layers, where each layer 
represents a distinct type of relation among the same set of nodes. 
They define a multi-layer logic in which the joint distribution of 
the layers is expressed through a sequence of conditional models, each describing 
the evolution from one layer to the next. Specifically, layer $k$ is modelled 
conditionally on the configuration of the preceding layer $k-1$, allowing for 
dependence across layers to be captured via sufficient statistics that encode 
edge persistence, reinforcement, or transition patterns. 
This formulation preserves the model flexibility for representing 
within-layer dependence while introducing a Markov-type dependence structure across layers. 

\section{Signed weighted networks}\label{sec:signed_nets}

Many real-world systems exhibit complex forms of interactions that cannot be adequately represented by binary relations alone. In numerous applications (such as gene regulatory networks, and biological or ecological interaction networks, social relations involving both cooperation and conflict) edges may carry values (weights) and signs (positive or negative), encoding both the strength and the polarity of interactions between nodes. 

A recent contribution in the ERGM literature for signed networks is by \cite{fri:meh:thu:kau22}. In this framework, the dyadic entries of the adjacency matrix $\bfY$ take values in $\{-1, 0, +1\}$, and the model adopts a categorical exponential-family form in the cross-sectional case and, in the dynamic setting, extends this with a first-order Markov process. The sufficient statistics $\bfs(\cdot)$ include counts of positive and negative edges, degree distributions by sign type, and, importantly, triadic closure patterns derived from structural balance theory. They employ geometrically weighted specifications of shared-partner statistics to reduce the risk of degeneracy \citep{sch11}.

Extending ERGMs to signed weighted networks allows for a more informative probabilistic representation of such complex relational structures. In a signed weighted network, each dyad $Y_{ij}$ can take positive and negative real (or integer) values. The distribution of $\bfY$ can then be modelled within the exponential-family framework by specifying sufficient statistics $s(\bfy)$ that capture both topological and weighted features. However, directly modelling signed weights is difficult because the presence, sign, and strength of edges are jointly dependent. This becomes particularly demanding when edge weights are continuous or integer-valued, complicating both the computation and interpretation of signed weighted ERGMs.

In this paper, we focus on polytomous networks \citep{pattison1999logit} and adopt a multilayer dissolution stochastic process \citep{cai:gol20} that represents the weighted network structure through a sequence of layer-specific transition dynamics.

Our modelling approach to signed networks follows the framework of \citet{ler16}, which represents a signed network by decomposing the overall generative process $\mathbf{y}$ into two components: an interaction structure that determines which dyads are non-zero, and a conditional weighted signed process that governs both the magnitude and the sign of those interactions.

Formally, let $\mathbf{y}$ be an observed signed, weighted adjacency matrix. We introduce a sequence of binary network layers $\mathbf{x}_{0:K} = \{\mathbf{x}_0, \mathbf{x}_1, \mathbf{x}_2, \ldots, \mathbf{x}_K\}$, where $\mathbf{x}_0$ indicates the presence of any interaction between nodes, and each higher layer $\mathbf{x}_k$ for $k>0$ defines a nested subset of progressively stronger interactions, with $X_{k,ij}=1$ if $Y_{ij} \ge k.$ The sign of each dyad present in $\mathbf{x}_0$ is encoded separately in a binary adjacency matrix $\mathbf{z}$. For convenience, we write $\mathbf{y}_k = \{\mathbf{z}, \mathbf{x}_k\}$. For example, if dyad $(i,j)$ exhibits a negative interaction at the strength level $k=2$, then $Y_{2,ij} = -1 \Leftrightarrow \{X_{1,ij}=1,\; X_{2,ij}=1,\; Z_{ij}=-1\}.$ This transformation allows the complex signed weighted network to be represented as a sequence of nested binary layers, each capturing interactions of a given intensity, while preserving the sign information. 

\section{Multi-layer dissolution models}\label{sec:multi_layer}

The multi-layer dissolution ERGM approach of \cite{cai:gol20} offers a robust and flexible method for modelling weighted networks, 
which are common in real-world applications but challenging for traditional binary ERGMs.
Multi-layer dissolution ERGMs are constructed by decomposing the weighted network into multiple binary layers, 
each defined according to the thresholds outlined in the previous sections.
This transformation allows the model to leverage the rich and thoroughly explored features of binary ERGM, including well-defined and interpretable network statistics and various modelling variants designed to solve potential issues like degeneracy.
The multi-layer dissolution ERGM is a relatively computationally efficient approach as the the estimation speed is less than linearly scaled with the number of layers, as higher layers constrain the dyadic space, reducing the computational burden of the simulation steps.

We now assume that $\mathbf{x}_{0:K}$ capture information on the evolution of edge weights across nested binary layers regardless of their sign structure. 
Each successive layer $k$ represents an incremental level of interaction strength or edge persistence, so that higher layers correspond to increasingly stable or intense relations. 
We adopt a multi-layer dissolution process to jointly model the signed and weighted network structure. 
Under this specification, each layer $\mathbf{x}_k$ and the sign structure $\bfz$ are jointly modelled conditionally on the observed preceding layer $\tilde{\mathbf{x}}_{k-1}$, which defines the support of $\mathbf{x}_k$ and allows edges to dissolve or persist as the interaction-strength threshold increases.
Formally, we define:

\begin{equation}\label{eq:psep_ML}
p(\mathbf{z}, \mathbf{x}_{1:K} \mid \mathbf{x}_0, \tilde{\mathbf{x}}_{1:K-1}, \bfvarphi_{1:K})
= \frac{1}{\kappa(\bfvarphi_{1:K})}
\exp\left\{\sum_{k=1}^K \bfvarphi_{k}^\top \mathbf{s}(\mathbf{z}, \mathbf{x}_k)\right\}
\prod_{k=1}^{K} \mathds{1}\{\mathbf{x}_k \subseteq \tilde{\mathbf{x}}_{k-1}\},
\end{equation}
where $\mathbf{s}(\mathbf{z}, \mathbf{x}_k)$ denotes the vector of sufficient statistics of layer $k$, computed on the edges of $\mathbf{x}_k$ and their signs, $\tilde{\mathbf{x}}_{k}$ denotes the observed $k$-th layer (with $\tilde{\mathbf{x}}_0 = \mathbf{x}_0$), $\mathbf{z} \in \{-1, +1\}^{|\mathbf{x}_0|}$ assigns a sign to every edge of the baseline layer, and $\kappa(\bfvarphi_{1:K})$ is the normalising constant, obtained by summing over all signs $\mathbf{z}$ and all layers $\mathbf{x}_1, \ldots, \mathbf{x}_K$ satisfying the support constraints. The transition from the observed layer $\tilde{\mathbf{x}}_{k-1}$ to $\mathbf{x}_k$ enters only through the support constraint.

The support of each layer $\mathbf{x}_k$ is the observed preceding layer
$\tilde{\mathbf{x}}_{k-1}$. The model is
therefore conditional on the observed lower layers: for each transition it describes which of
the edges observed at level $k-1$ persist at level $k$, and with which sign, given everything
observed below. Two consequences follow. First, the nesting
$\mathbf{x}_K \subseteq \cdots \subseteq \mathbf{x}_1 \subseteq \mathbf{x}_0$ holds for the observed
data but is not imposed on the joint distribution: a network drawn from
Equation~\eqref{eq:psep_ML} need not have nested layers, and hence need not correspond to a
valid weighted network. This affects simulation-based summaries (auxiliary networks in the
exchange algorithm and predictive checks), which should be read transition by transition, but
not the likelihood of the observed data. Second, because the sign structure $\mathbf{z}$ is
shared by all layers, the transitions are not conditionally independent and the model has a
single normalising constant; the layer-specific parameters are therefore estimated jointly
rather than from $K$ separate models.

A fully generative alternative would allow each layer to depend on the random preceding layer, replacing $\tilde{\mathbf{x}}_{k-1}$ with $\mathbf{x}_{k-1}$ in the support constraints. This would define a joint distribution over nested signed weighted networks, with a normalising constant accounting for the dependence between layers. We adopt the conditional formulation instead, which is consistent with the dissolution interpretation of the layers and retains support constraints determined by the preceding observed layer. The computational methods developed in this paper could, however, be readily adapted to the fully generative formulation.

While the multi-layer dissolution process in Equation~\ref{eq:psep_ML} assumes a first-order Markov dependence, the framework can be naturally extended to allow higher-order dependencies. 

\subsection{Network statistics}\label{mod_specs}

The vector of binary sufficient statistics in Equation~\ref{eq:psep_ML} can incorporate a broad range of network features that characterise the persistence properties of the dissolution process \citep{sni:pat:rob:han06}. For example, they may include statistics capturing the density by counting the number of positive (\texttt{edges+}) or negative (\texttt{edges-}) edges.

\begin{figure}[htbp]
   \centering
   \includegraphics[scale = 0.33]{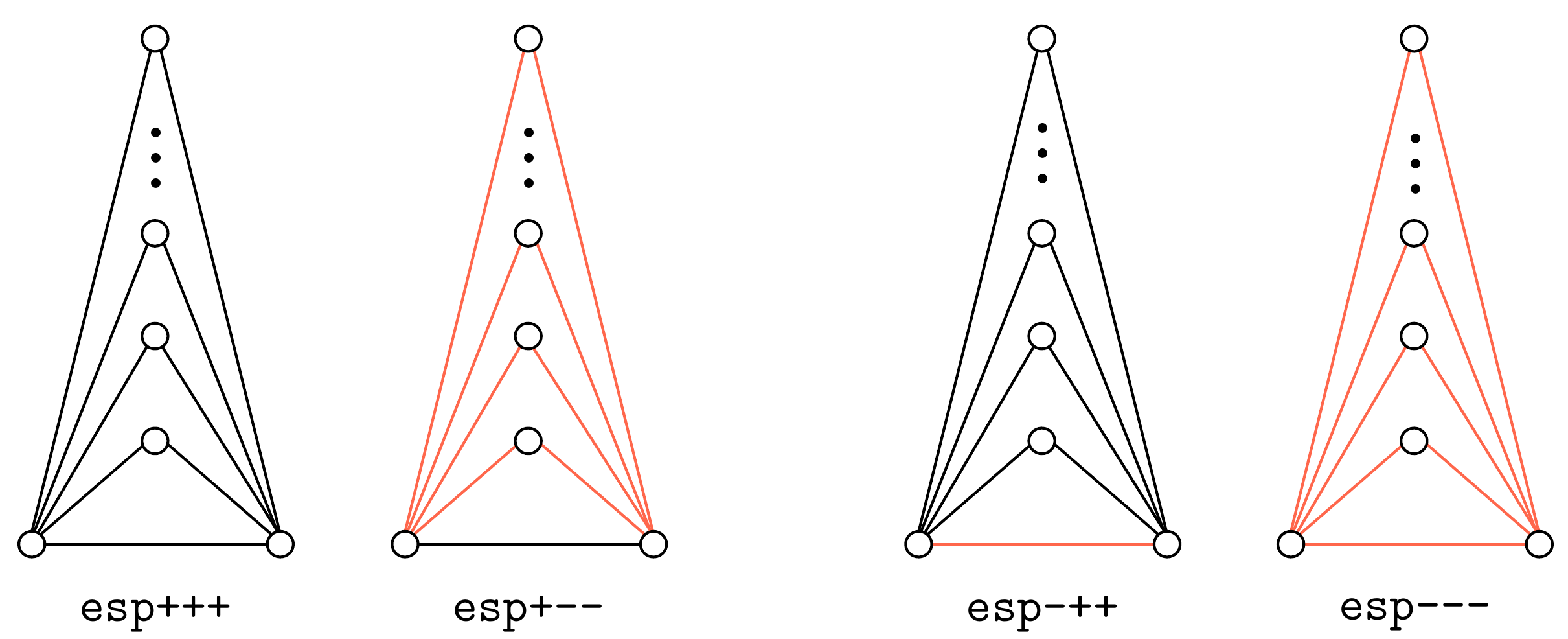}
   \caption{\texttt{esp+++} and \texttt{esp+--} represent balanced triads, respectively capturing the principle that ``the friends of my friends 
   are friends'' and ``the enemies of my friend are enemies.'' In contrast, \texttt{esp-++} and \texttt{esp---} represent unbalanced triads, corresponding 
   respectively to ``the friends of my enemies are my friends'' and ``the enemies of my enemy are my enemies''.}
   \label{fig:sbt_triads}
\end{figure}

Triadic configurations used to assess structural balance, shown in Figure \ref{fig:sbt_triads}, can be expressed through geometrically weighted functions of edgewise shared partners (ESP). Specifically, these include geometrically weighted functions of ESP configurations with positive base and positive two-paths (\texttt{gwesp+++}); with positive base and negative two-paths (\texttt{gwesp+--}); with negative base and positive two-paths (\texttt{gwesp-++}); and with negative base and negative two-paths (\texttt{gwesp---}). Equivalent configurations were implemented by \cite{fri:meh:thu:kau22} under different terminology.

Exogenous network statistics may include positive or negative tie density among nodes with the same categorical covariate value, as well as functions of covariate dissimilarity, such as the absolute difference between non-categorical covariates for a pair of nodes.

\subsection{Model interpretation}\label{mod_interpr}

We now derive a local dyadic-level interpretation of the multi-layer dissolution model defined in Equation~\ref{eq:psep_ML}. 

The conditional probability of observing a positive edge in dyad $(i,j)$ at layer $k,$ denoted by 
$Y_{k,ij} = +1 \Leftrightarrow \{Z_{ij} = +1, X_{k,ij} = 1\},$ follows a categorical distribution:
$$
\Pr\big(Y_{k,ij}= +1 \mid \tilde{X}_{k-1,ij} = 1, \bfy_{k,(-ij)}, \tilde{\bfx}_{0:k-1}\big) = 
\frac{\exp\left\{\bfvarphi_k^\top\bfs\left(\bfy_{k,ij}^{(+1)}\right)\right\}}
{\sum_{y \in \{-1, 0, +1\}}\exp\left\{\bfvarphi_k^\top\bfs\left(\mathbf{y}_{k,ij}^{(y)}\right)\right\}},
$$
where $\bfy_{k,(-ij)}$ is the observed signed layer structure $k$ in which the dyad $(i,j)$ is excluded and $\mathbf{y}_{k,ij}^{(+1)}$ and $\mathbf{y}_{k,ij}^{(y)}$ denote the binary signed network configuration in which the dyad $(i,j)$ is set to $+1$ or to $y \in \{-1,0,+1\}$, respectively, while all other dyads remain fixed. Here $\tilde{X}_{k-1,ij} = 1$ indicates that dyad $(i,j)$ belongs to the observed layer $k-1$, so that it can be active at layer $k$; for dyads outside $\tilde{\bfx}_{k-1}$, $Y_{k,ij} = 0$ with probability one. This formulation allows the signed network layer $\bfz$ to be influenced by transitions in other layers, enabling dependencies across layers to be incorporated directly into the conditional distribution.

The conditional log-odds of observing $Y_{k,ij} = +1$ relative to $Y_{k,ij} = 0$ are given by:
$$
\log \left(\frac{\Pr(Y_{k,ij} = +1 \mid Y_{k-1,ij} = +1, \bfy_{k,(-ij)}, \tilde{\bfx}_{0:k-1})}
                {\Pr(Y_{k,ij} = 0 \mid Y_{k-1,ij}=+1, \bfy_{k,(-ij)}, \tilde{\bfx}_{0:k-1})} \right)
= \bfvarphi_k^\top \Delta_{k,ij}^{0 \rightarrow +1} \left( \bfy_{k,ij} \right),
$$
where $\Delta_{k,ij}^{0 \rightarrow +1} \left( \bfy_{k,ij} \right)$ denotes the change statistic obtained by toggling dyad at layer $k$ from $0$ to $+1,$ i.e., $\bfs\left(\bfy_{k,ij}^{(+1)}\right) - \bfs\left(\bfy_{k,ij}^{(0)}\right).$ Analogously, log-odds of observing $Y_{k,ij} = -1$ relative to $Y_{k,ij} = 0$ are:
\begin{equation*}
\log \left(\frac{\Pr(Y_{k,ij} = -1 \mid Y_{k-1,ij} = -1, \bfy_{k,(-ij)}, \tilde{\bfx}_{0:k-1})}
                {\Pr(Y_{k,ij} = 0 \mid Y_{k-1,ij}= -1, \bfy_{k,(-ij)}, \tilde{\bfx}_{0:k-1})} \right)
= \bfvarphi_k^\top \Delta_{k,ij}^{0 \rightarrow -1} \left( \bfy_{k,ij} \right).
\end{equation*}
Note that the sign of $Y_{k,ij}$ must match that of $Y_{k-1,ij}$ or be $0$ as the sign cannot switch across layers.

The conditional log-odds of observing $Z_{ij} = +1$ relative to $Z_{ij} = -1$ are:
$$
\log\left(
\frac{\Pr(Z_{ij}= +1 \mid \bfz_{-(ij)}, \bfy_{1:K,-(ij)}, \tilde{\bfx}_{0:K})}
     {\Pr(Z_{ij}= -1 \mid \bfz_{-(ij)}, \bfy_{1:K,-(ij)}, \tilde{\bfx}_{0:K})}
    \right)
=
\sum_{k: x_{k,ij}=1} \bfvarphi_k^\top
\Delta_{k,ij}^{-1 \rightarrow +1} \left( \bfy_{k,ij} \right),
$$
where $\Delta_{k,ij}^{-1 \rightarrow +1} \left( \bfy_{k,ij} \right)$ 
denotes the change statistic obtained by flipping the sign of dyad $(i,j)$
from $-1$ to $+1$ in all layers $k$ such that $X_{k,ij}=1$.
This quantity represents the cumulative information across layers favouring a 
positive over a negative relation between $i$ and $j.$ Layers where $X_{k,ij}=1$ contribute 
to this information, while inactive layers $(X_{k,ij}=0)$ do not affect the sign inference. 
Intuitively, a dyad will tend to have $Z_{ij}=+1$ when forming or maintaining a positive 
relation consistently increases the model sufficient statistics across layers.

\section{Bayesian inference}\label{sec:Bayes}

To allow for heterogeneity across layers while retaining a shared dependence structure, 
we specify a hierarchical prior on the layer-specific ERGM parameters 
$\boldsymbol{\varphi}_k$. 
Each $\boldsymbol{\varphi}_k$ governs the network formation process at layer $k$, 
and the hierarchical specification enables partial pooling across layers. 
Formally, we assume:
$$
\begin{aligned}
\boldsymbol{\mu}_{\boldsymbol{\varphi}} &\sim \mathcal{N}(\boldsymbol{\mu}_{\boldsymbol{\varphi}_0}, \boldsymbol{\Sigma}_{\boldsymbol{\varphi}_0}), \\
\boldsymbol{\Sigma}_{\boldsymbol{\varphi}} &\sim \mathcal{W}^{-1}(\nu_{\boldsymbol{\varphi}_0}, \mathbf{S}_{\boldsymbol{\varphi}_0}), \\
\boldsymbol{\varphi}_k \mid \boldsymbol{\mu}_{\boldsymbol{\varphi}}, \boldsymbol{\Sigma}_{\boldsymbol{\varphi}} &\overset{iid}{\sim} 
\mathcal{N}(\boldsymbol{\mu}_{\boldsymbol{\varphi}}, \boldsymbol{\Sigma}_{\boldsymbol{\varphi}}),
\end{aligned}
$$
where $\boldsymbol{\mu}_{\boldsymbol{\varphi}}$ represents the overall mean vector 
and $\boldsymbol{\Sigma}_{\boldsymbol{\varphi}}$ the covariance structure governing 
the variability of the layer-specific parameters. 
This hierarchical prior allows information to be shared across layers, 
shrinking the estimates of $\boldsymbol{\varphi}_k$ toward a common mean 
when layers exhibit similar structural features, 
while still allowing for layer-specific deviations.

The layer-specific parameter vectors $\boldsymbol{\varphi}_k$ capture the strength and direction of dependence associated with the network statistics $\mathbf{s}(\mathbf{z}, \mathbf{x}_{k})$, and therefore govern which edges of the observed layer $\tilde{\mathbf{x}}_{k-1}$ persist at layer $k$, and with which sign. 
Positive values of a parameter $\varphi_{k,r}$ indicate that the corresponding configuration (e.g., signed density, balanced triadic closure) increases the likelihood of observing an edge pattern that contributes to the statistic $s_r(\cdot)$, whereas negative coefficients penalise such configurations. 
The mean hyperparameter $\boldsymbol{\mu}_{\boldsymbol{\varphi}}$ represents the overall, population-level average effect across layers, summarising the common structural features that persist throughout the multi-layer structure. 
The covariance hyperparameter $\boldsymbol{\Sigma}_{\boldsymbol{\varphi}}$ quantifies the degree of heterogeneity across layers and the correlation between different structural mechanisms, thereby capturing how the importance of various network effects evolves across layers. 

\subsection{Posterior estimation}

To sample from the posterior distribution of $\boldsymbol{\varphi}_{1:K}, \boldsymbol{\mu}_{\boldsymbol{\varphi}}, 
\boldsymbol{\Sigma}_{\boldsymbol{\varphi}}$ we implement an 
adaptive approximate exchange 
algorithm which extends the one introduced by \citet{cai:fri11} to 
the hierarchical multi-layer dissolution setting, allowing joint inference 
on the layer-specific parameters $\boldsymbol{\varphi}_{k}$ and their hierarchical 
hyperparameters $(\boldsymbol{\mu}_{\varphi}, \boldsymbol{\Sigma}_{\varphi})$. 

Let $\bfvarphi = (\bfvarphi_1^\top, \ldots, \bfvarphi_K^\top)^\top$ collect the layer-specific parameters, and let $\bfs(\bfy)$ be the corresponding vector of sufficient statistics of the observed network $\bfy = (\mathbf{z}, \bfx_{1:K})$, obtained by stacking $\bfs(\mathbf{z}, \bfx_k)$ for $k = 1, \ldots, K$. The components of $\bfvarphi$ are partitioned into blocks, which are updated in turn by exchange steps. Two partitions are used in this paper: one block per layer (layer-wise update), so that $\bfvarphi_k$ is updated with the other layers held fixed, and a single block containing all $Kp$ parameters (block update). For block $b$, a proposal $\bfvarphi'$ is obtained by perturbing the components in the block, $\bfvarphi'_b \sim \mathcal{N}(\bfvarphi_b, \gamma_b^2 \mathbf{B}_b)$, and leaving the others unchanged, where $\mathbf{B}_b$ is a proposal covariance matrix and $\gamma_b$ is a scalar scale factor \citep{rob:ros09}.

Because the layers share the sign structure $\mathbf{z}$, the model in Equation~\ref{eq:psep_ML} has a single normalising constant. An auxiliary network $\bfy' = (\mathbf{z}', \bfx'_{1:K})$ is therefore simulated from the joint model $p(\cdot \mid \tilde{\bfx}_{0:K-1}, \bfvarphi')$ by running a Markov chain started at the observed network, and the proposal is accepted with probability $\min(1, \alpha)$, where
$$\log \alpha = (\bfvarphi - \bfvarphi')^\top \left[\bfs(\bfy') - \bfs(\bfy)\right] + \sum_{k \in b} \left\{\log \pi(\bfvarphi'_{k} \mid \bfmu_{\bfvarphi}, \bfSigma_{\bfvarphi}) - \log \pi(\bfvarphi_{k} \mid \bfmu_{\bfvarphi}, \bfSigma_{\bfvarphi})\right\},$$
which avoids evaluation of the intractable normalising constant \citep{Murray06, cai:fri11}; the sum runs over the layers whose parameters are in block $b$.

During an initial adaptation phase, the scales $\gamma_b$ are tuned toward a target acceptance rate, e.g., $a_{\text{target}} = 0.234$ \citep{li:etal25}, and, optionally, the covariance matrices $\mathbf{B}_b$ are re-estimated from the history of the chain, as in the adaptive Metropolis algorithm of \citet{haa:sak:tam01}. This matters when the posterior is strongly correlated, as happens for edge-count and triadic-closure statistics in dense layers (Section~\ref{sec:app}): an isotropic proposal must then take steps as small as the narrowest direction of the posterior and moves very slowly along the others. Adaptation is stopped at the end of the burn-in, so that the retained draws come from a Markov chain with a fixed transition kernel. After the exchange steps, the hyperparameters $(\bfmu_{\bfvarphi}, \bfSigma_{\bfvarphi})$ are sampled via conjugate Gibbs updates from their Normal and inverse-Wishart full conditional distributions. The complete procedure is given in Algorithm~\ref{alg:ae}.

\begin{algorithm}
\caption{Adaptive exchange for hierarchical multi-layer dissolution ERGMs.}
\label{alg:ae}
\begin{algorithmic}[1]
\begin{spacing}{1.2}
\State \textbf{Input:} Blocks $b = 1, \ldots, n_B$ of the components of $\bfvarphi$ (one block per layer, or a single block); initial values $\bfvarphi^{(0)}$, $\gamma_b$ and $\mathbf{B}_b^{(0)}$; adaptation interval $\Delta t_{\text{adapt}}$, start $t_{\text{start}}$ and stop $t_{\text{stop}}.$
\State \textbf{Initialise:} $\bfvarphi \gets \bfvarphi^{(0)}$; $\mathbf{B}_b \gets \mathbf{B}_b^{(0)}$; $\bfmu_{\bfvarphi} \gets \bfmu_{\bfvarphi_0}$; $\bfSigma_{\bfvarphi} \gets \mathbf{S}_{\bfvarphi_0}/(\nu_{\bfvarphi_0} - p - 1)$
\For{$t = 1$ to $T$}
    \State \textbf{Exchange update:}
    \For{each block $b$}
        \State $\bfvarphi' \gets \bfvarphi$; \quad $\bfvarphi'_b \sim \mathcal{N}(\bfvarphi_b, \gamma_b^2 \mathbf{B}_b)$
        \State $\bfy' \sim p(\cdot \mid \tilde{\bfx}_{0:K-1}, \bfvarphi')$        
        \State $\log \alpha \gets (\bfvarphi - \bfvarphi')^\top \left[\bfs(\bfy') - \bfs(\bfy)\right] + \sum_{k \in b} \log \dfrac{\pi(\bfvarphi'_{k} \mid \bfmu_{\bfvarphi}, \bfSigma_{\bfvarphi})}{\pi(\bfvarphi_{k} \mid \bfmu_{\bfvarphi}, \bfSigma_{\bfvarphi})}$
        \If{$\log u < \log \alpha$ where $u \sim \text{Uniform}(0,1)$}
            \State $\bfvarphi \gets \bfvarphi'$
        \EndIf
    \EndFor
    \State \textbf{Adaptation:}
    \If{$t_{\text{start}} \leq t \leq t_{\text{stop}} \land (t \bmod \Delta t_{\text{adapt}}) = 0$}
        \For{each block $b$}
            \State $\bar{a}_b \gets$ acceptance rate of block $b$ in the latest window
            \If{$\bar{a}_b > a_{\text{target}}$} for some $\lambda > 0$
                \State $\gamma_b \gets \gamma_b (1 + \lambda)$ 
            \Else
                \State $\gamma_b \gets \gamma_b (1 - \lambda)$
            \EndIf
            \State (optional) $\mathbf{B}_b \gets 0.95\, \widehat{\mathrm{Cov}}\big(\bfvarphi_b^{(\lceil t/2 \rceil)}, \ldots, \bfvarphi_b^{(t)}\big) + 0.05\, \mathbf{B}_b^{(0)}$
        \EndFor
    \EndIf
    \State \textbf{Gibbs update:}
    \State $\bfSigma_{\mu} \gets \left( \bfSigma_{\bfvarphi_0}^{-1} + K \bfSigma_{\bfvarphi}^{-1} \right)^{-1}$; \quad $\bar{\bfvarphi} \gets \frac{1}{K} \sum_k \bfvarphi_k$
    \State $\bfmu_{\bfvarphi} \sim \mathcal{N}\left(\bfSigma_{\mu}(\bfSigma_{\bfvarphi_0}^{-1} \bfmu_{\bfvarphi_0} + K \bfSigma_{\bfvarphi}^{-1} \bar{\bfvarphi}), \bfSigma_{\mu}\right)$
    \State $\bfSigma_{\bfvarphi} \sim \mathcal{W}^{-1}\big(\nu_{\bfvarphi_0} + K,\ \mathbf{S}_{\bfvarphi_0} + \sum_k (\bfvarphi_k - \bfmu_{\bfvarphi})(\bfvarphi_k - \bfmu_{\bfvarphi})^\top\big)$
\EndFor
\end{spacing}
\end{algorithmic}
\end{algorithm}

\section{Simulation study}\label{sec:sim}

We conducted a simulation study to assess whether the adaptive exchange algorithm of Section~\ref{sec:Bayes} recovers known layer-specific parameters, and to separate the behaviour of the sampler from the shrinkage induced by the hierarchical prior. The design reproduces the size and density of the network analysed in Section~\ref{sec:app}, so that the results apply to the setting in which the method is used there.

\subsection{Design and model specification}\label{sec:sim_design}

The baseline layer $\mathbf{x}_0$ was a single Bernoulli random graph on $n = 78$ nodes with edge probability $0.69$, giving $2{,}081$ edges (density $0.693$); it was held fixed across replications.

We specify a multi-layer dissolution ERGM incorporating four endogenous statistics in each layer,
capturing both local connectivity and higher-order dependencies among signed edges:
\begin{itemize}
\item $\texttt{edges+},$ the number of positive edges, representing overall network density;
\item $\texttt{gwesp+++}(\alpha)$: the geometrically weighted count of positive edges by their number of shared partners connected to both endpoints by positive edges, capturing balanced closure among positive interactions;
\item $\texttt{gwesp+--}(\alpha)$: the geometrically weighted count of positive edges by their number of shared partners connected to both endpoints by negative edges;
\item $\texttt{gwesp-++}(\alpha)$: the geometrically weighted count of negative edges by their number of shared partners connected to both endpoints by positive edges.
\end{itemize}
All statistics of the $k$-th transition are computed on layer $\bfx_k$ and on the signs $\mathbf{z}$ of its edges. The decay parameters $\alpha$ of the geometrically weighted statistics are set to $\alpha_1 = 0.4$, $\alpha_2 = 0.6$, and $\alpha_3 = 2$ for the first, second and third transitions, respectively. With a fixed decay, an edge with $m$ shared partners contributes $e^{\alpha}\{1 - (1 - e^{-\alpha})^{m}\}$ to the statistic. Small values of $\alpha$ make this contribution approach its maximum $e^{\alpha}$ after very few shared partners, so the statistic mainly distinguishes edges that are embedded in at least one triangle from edges that are not; larger values continue to reward additional shared partners. Small decays are therefore used in the dense lower layers, where most positive edges have many shared partners  and a slowly saturating statistic would be close to degeneracy, and a larger decay in the sparse top layer, where triangles are few and their number is informative. A consequence of this choice is that, in the first two transitions, \texttt{gwesp+++} is close to $e^{\alpha_k}$ times \texttt{edges+}, and the two parameters are identified separately only through the positive edges with few positive shared partners.

Each dataset was simulated by MCMC from the model at the true value $\bfvarphi^\star$ (reported in Table~\ref{tab:sim}), starting from a random configuration compatible with the supports and discarding the first $100{,}000$ steps.

\subsection{Identifiability check} 

Before running the study we verified that the simulated data are informative about every parameter at the true values. For an exponential-family model the covariance matrix of the sufficient statistics, $\mathcal{I}(\bfvarphi^\star) = \mathrm{Cov}_{\bfvarphi^\star}\{\mathbf{s}\}$, is the Fisher information, and the square roots of the diagonal of $\mathcal{I}(\bfvarphi^\star)^{-1}$ are the standard errors implied by the likelihood alone. Over $2{,}000$ networks simulated at $\bfvarphi^\star$ ($1{,}000$ on each of two independent draws of the supports), every statistic was non-zero in at least $98\%$ of the networks and the likelihood-only standard errors ranged from $0.09$ to $0.47$. This check proved necessary as signed triadic effects can be recovered only from networks that contain enough triangles of the corresponding type.

\subsection{Algorithm settings} 

Algorithm~\ref{alg:ae} was run with the layer-wise update for $8{,}000$ iterations from $\bfvarphi_k \sim \mathcal{N}(\mathbf{0}, 0.01\,\mathbf{I}_4)$, with fixed proposal covariance $\mathbf{B}_k = 0.05\,\mathbf{I}_4$ and initial scales $\gamma_k = 0.45$, $0.30$ and $0.15$ for $k = 1, 2, 3$. Each auxiliary network was obtained by $20{,}000$ MCMC steps started at the observed network. The proposal scales were adapted every $250$ iterations between iterations $250$ and $2{,}000$, using the acceptance rate over the preceding $250$ iterations, $\lambda = 0.2$ and $a_{\text{target}} = 0.234$, and were then held fixed, so that the retained draws come from a Markov chain with a fixed transition kernel. The first $2{,}000$ iterations were discarded as burn-in, leaving $6{,}000$ draws per chain. To reduce computation, the model and proposal objects used for the auxiliary simulations were constructed once per chain and reused at every exchange step, so that each step only incurs the cost of the auxiliary MCMC itself. The procedure was repeated for $20$ independently simulated datasets.

\subsection{Results}\label{sec:sim_results}

We used the hierarchical prior of Section~\ref{sec:Bayes} with $\bfmu_{\bfvarphi_0} = \mathbf{0}$, $\bfSigma_{\bfvarphi_0} = 3\,\mathbf{I}_4$, $\nu_{\bfvarphi_0} = 15$ and $\mathbf{S}_{\bfvarphi_0} = \mathbf{I}_4$. For each parameter we recorded the posterior mean and the equal-tailed 95\% credible interval, and computed across replications the bias of the posterior mean with its Monte Carlo standard error (MCSE), the root mean squared error (RMSE), and the coverage and average width of the 95\% credible intervals.

Table~\ref{tab:sim} summarises the posterior results. The posterior means are close to the true values for all twelve parameters: the absolute bias is at most $0.111$ (median $0.039$) and never exceeds one fifth of the average interval width, and the RMSE ranges from $0.095$ to $0.250$. The coverage of the 95\% credible intervals ranges from $0.85$ to $1.00$, with $11$ of the $12$ parameters at or above the nominal level. Across replications, the standard deviation of the posterior means is between $0.68$ and $1.09$ times the average posterior standard deviation implied by the interval widths, so the credible intervals are not too narrow. The mean acceptance rates of the exchange updates were $0.24$, $0.25$ and $0.33$ for the three layers.

\begin{table}[htbp]
\centering
\caption{Simulation results over $20$ replications. For each layer transition and statistic: true value ($\bfvarphi^\star_k$); bias of the posterior mean with its Monte Carlo standard error (MCSE); root mean squared error (RMSE); coverage and average width of the 95\% credible intervals. The last two columns report the bias and interval width implied by the hierarchical prior alone, under a Normal approximation to the likelihood.}\label{tab:sim}
\small
\setlength{\tabcolsep}{5pt}
\begin{tabular}{c|l|r|rrrr|rr}
\hline
 & & & \multicolumn{4}{c|}{Adaptive exchange} & \multicolumn{2}{c}{Prior-implied} \\
Layer & Parameter & $\bfvarphi^\star_k$ & Bias (MCSE) & RMSE & Coverage & Width & Bias & Width \\
\hline
\multirow{4}{*}{1}
 & $\varphi_{1,  \texttt{edges+}}$      & $0.62$  & $-0.111$ ($0.027$) & $0.161$ & $0.95$ & $0.56$ & $-0.078$ & $0.53$ \\
 & $\varphi_{1,  \texttt{gwesp+++}}$  & $-0.50$ & $-0.017$ ($0.024$) & $0.107$ & $0.95$ & $0.39$ & $0.044$  & $0.34$ \\
 & $\varphi_{1,  \texttt{gwesp+--}}$   & $0.20$  & $0.022$ ($0.024$)  & $0.108$ & $0.95$ & $0.43$ & $-0.001$ & $0.40$ \\
 & $\varphi_{1,  \texttt{gwesp-++}}$  & $-1.00$ & $-0.007$ ($0.022$) & $0.095$ & $1.00$ & $0.38$ & $0.025$  & $0.36$ \\
\hline
\multirow{4}{*}{2}
 & $\varphi_{2,  \texttt{edges+}}$      & $-0.20$ & $0.046$ ($0.024$)  & $0.114$ & $1.00$ & $0.49$ & $0.057$  & $0.48$ \\
 & $\varphi_{2,  \texttt{gwesp+++}}$  & $0.20$  & $-0.080$ ($0.027$) & $0.142$ & $0.85$ & $0.43$ & $-0.049$ & $0.40$ \\
 & $\varphi_{2,  \texttt{gwesp+--}}$   & $-0.25$ & $-0.001$ ($0.024$) & $0.105$ & $1.00$ & $0.43$ & $0.018$  & $0.43$ \\
 & $\varphi_{2,  \texttt{gwesp-++}}$  & $-1.30$ & $0.032$ ($0.034$)  & $0.152$ & $1.00$ & $0.88$ & $0.065$  & $0.92$ \\
\hline
\multirow{4}{*}{3}
 & $\varphi_{3,  \texttt{edges+}}$      & $0.08$  & $0.004$ ($0.027$)  & $0.119$ & $0.95$ & $0.49$ & $0.030$  & $0.48$ \\
 & $\varphi_{3,  \texttt{gwesp+++}}$  & $0.20$  & $-0.076$ ($0.029$) & $0.147$ & $1.00$ & $0.58$ & $-0.045$ & $0.51$ \\
 & $\varphi_{3,  \texttt{gwesp+--}}$   & $-0.15$ & $-0.063$ ($0.036$) & $0.167$ & $1.00$ & $0.76$ & $0.030$  & $0.74$ \\
 & $\varphi_{3,  \texttt{gwesp-++}}$  & $-1.02$ & $-0.087$ ($0.054$) & $0.250$ & $1.00$ & $1.16$ & $-0.042$ & $1.15$ \\
\hline
\end{tabular}
\end{table}

Three parameters show a bias exceeding twice its MCSE: \texttt{edges+} in the first transition ($-0.111$, MCSE $0.027$) and \texttt{gwesp+++} in the second ($-0.080$, MCSE $0.027$) and third ($-0.076$, MCSE $0.029$); \texttt{gwesp+++} in the second transition also has the lowest coverage ($0.85$). All three are shifts towards the mean across layers, as expected under the hierarchical prior. With $\nu_{\bfvarphi_0} = 15$ and $\mathbf{S}_{\bfvarphi_0} = \mathbf{I}_4$, the prior expectation of $\bfSigma_{\bfvarphi}$ is $\mathbf{S}_{\bfvarphi_0}/(\nu_{\bfvarphi_0} - p - 1) = 0.1\,\mathbf{I}_4$, a between-layer standard deviation of about $0.32$, and with only $K = 3$ layers $15$ of the $18$ degrees of freedom of the conditional posterior of $\bfSigma_{\bfvarphi}$ come from the prior. The true \texttt{edges+} and \texttt{gwesp+++} values vary more across layers than this (standard deviations $0.42$ and $0.40$, against $0.24$ for \texttt{gwesp+--} and $0.17$ for \texttt{gwesp-++}), so their layer-specific estimates are partially pooled towards the common mean.

To quantify this effect independently of the exchange algorithm, we computed the bias, coverage and interval width implied by the prior alone. We approximated the sampling distribution of the maximum likelihood estimator by $\hat{\bfvarphi} \sim \mathcal{N}(\bfvarphi^\star, \mathcal{I}(\bfvarphi^\star)^{-1})$, with $\mathcal{I}(\bfvarphi^\star)$ estimated from the networks simulated for the identifiability check, combined the corresponding Normal likelihood with the same hierarchical prior through an exact conjugate Gibbs sampler, and averaged over $400$ draws of $\hat{\bfvarphi}$. No network is simulated at this stage and no exchange step is involved, so any bias obtained in this way is due to the prior. The prior-implied bias and width, reported in the last two columns of Table~\ref{tab:sim}, reproduce the observed pattern: $-0.078$ for \texttt{edges+} in the first transition and $-0.049$ and $-0.045$ for \texttt{gwesp+++} in the second and third, with a prior-implied coverage of $0.92$ for \texttt{gwesp+++} in the second transition. On the basis of the values reported in Table~\ref{tab:sim}, the observed interval widths are within $8\%$ of the prior-implied widths for ten of the twelve parameters and $14$--$15\%$ wider for \texttt{gwesp+++} in the first and third transitions, indicating that the exchange algorithm explores the posterior distribution correctly, if anything slightly conservatively for these two parameters. The observed estimates lie on average $0.03$ below the prior-implied values; the difference exceeds twice its Monte Carlo error for two parameters only (\texttt{gwesp+++} in the first transition and \texttt{gwesp+--} in the third) and may reflect the Normal approximation, the finite length of the auxiliary chains or Monte Carlo variability.

With $20$ replications the MCSE of a coverage of $0.95$ is about $0.05$, so differences in coverage of that size cannot be resolved, whereas the MCSE of the bias ranges from $0.022$ to $0.054$. Overall, once the design contains enough triangles for each signed triadic statistic, the algorithm recovers the layer-specific parameters with small error and well-calibrated intervals, and the remaining bias is the partial pooling implied by the hierarchical prior.

\section{Application to gene co-expressions}\label{sec:app}

\subsection{Research question}

Dilated cardiomyopathy (DCM) is a genetically heterogeneous form of
heart failure whose associated genes span multiple functionally distinct
modules, with co-expression relationships that vary in both polarity and
magnitude. This biological complexity generates a network structure in
which signed triadic configurations, confidence-stratified interaction
evidence, and structural balance effects carry interpretable biological
meaning, making DCM a compelling test case for the proposed framework.
The availability of curated gene-disease associations, protein--protein
interaction data with multi-level reliability scores, and genome-wide
co-expression measurements provides the empirical foundation for a
multi-layer signed weighted network representation.

\subsubsection{Network construction and model specification}

Gene sets associated with Dilated cardiomyopathy (DOID:12930) were
retrieved from the Jensen Diseases database \citep{pletscher:2015}. Based on these
genes, a protein–protein interaction network was constructed using STRING (Homo sapiens,
version 12.0), integrating evidence from experiments, curated databases, co-expression,
gene neighborhood, and co-occurrence \citep{szklarczyk:2023}. The resulting interaction
scores were used to define an undirected binary network by applying a minimum confidence
threshold of 0.15, and including up to 30 first-shell interactors in order to maximize
sensitivity and capture an extended functional context around disease-associated genes.

To complement interaction data with functional relationships, gene
co-expression information was incorporated using the ARCHS4 resource
\citep{lachmann:2018}. Specifically, a genome-wide gene–gene correlation matrix
(\texttt{human\_correlation\_archs4.f}, version 1.0, July 2020), containing precomputed
pairwise Pearson correlation coefficients derived from large-scale RNA-seq samples, was
used. The matrix was subset to retain only genes present in the STRING-derived network;
two genes (\emph{CAVIN4} and \emph{KBTBD13}) were not represented in the ARCHS4
resource and were therefore excluded, yielding a final set of 78 genes. The sign of each edge, encoded in $\mathbf{z}$, was set to the sign of the corresponding ARCHS4 correlation coefficient. In addition, the network includes node-level covariates derived from the Hallmark gene sets collection of the Molecular Signatures Database (MSigDB) \citep{liberzon:2015}. 

The model includes the same endogenous statistics as in the simulation study, with the same decay parameters, together with the following two covariate-based statistics: 
\begin{itemize}
\item $\texttt{nm(myo)}$, the number of positive edges between genes sharing the same Myogenesis membership indicator, capturing the tendency of genes within (or outside) the Hallmark Myogenesis gene set to interact positively with one another;
\item $\texttt{nm(hall)}$, the number of positive edges between genes annotated to the same number of Hallmark gene sets, capturing homophily with respect to functional pleiotropy, i.e., the tendency of genes with similar breadth of biological involvement to be positively co-expressed.
\end{itemize}

\begin{figure}[htp]
   \centering
   \includegraphics[width = \textwidth]{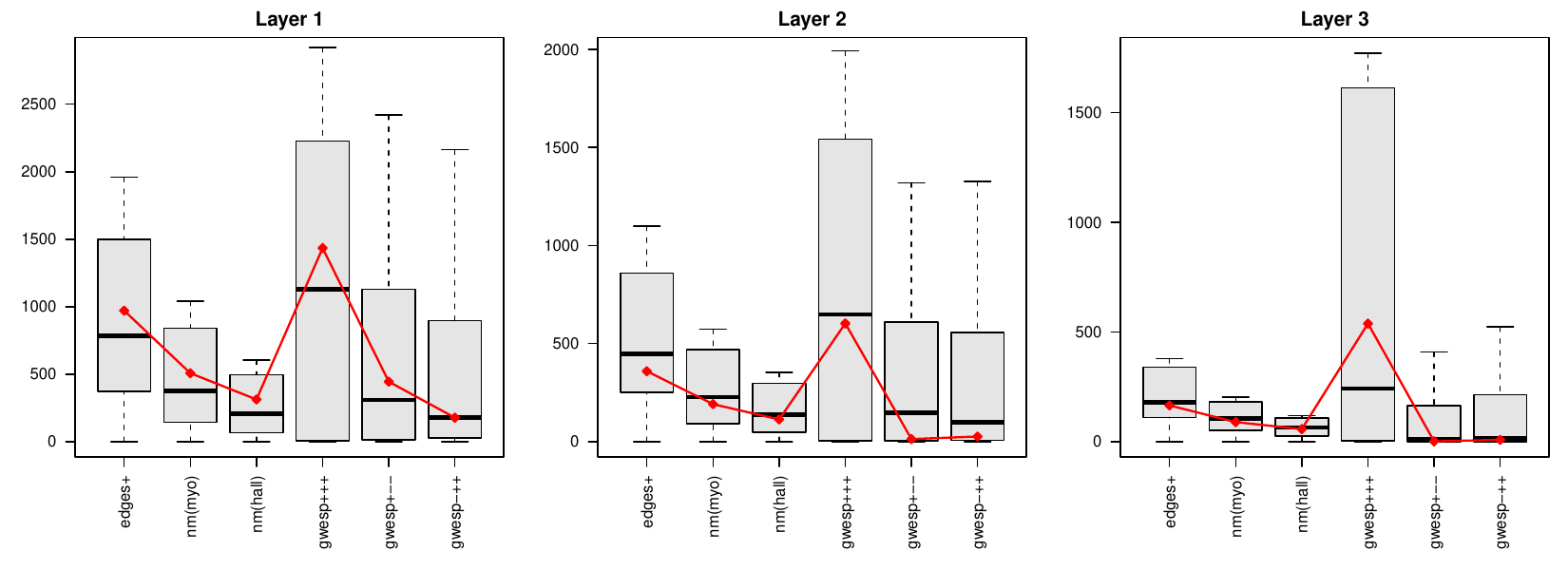}
   \caption{Prior predictive distributions of the six statistics of each layer transition, based on $1{,}000$ simulated networks and summarised by boxplots (outlying values not shown), with the observed values indicated by red diamonds connected by a solid red line.}
   \label{fig:prior_pred}
\end{figure}

We define the following hyper-priors: 
$\boldsymbol{\mu}_{\boldsymbol{\varphi}_0} = \mathbf{0}, \boldsymbol{\Sigma}_{\boldsymbol{\varphi}_0} = 3 \mathbf{I}_{6}, \nu_{\boldsymbol{\varphi}_0} = 15, \mathbf{S}_{\boldsymbol{\varphi}_0} = \mathbf{I}_{6}.$
To assess the prior specification, we conducted prior predictive checks. We drew $1{,}000$ parameter vectors from the hierarchical prior, simulated one network for each by running $50{,}000$ MCMC steps from the observed network, and computed the statistics included in the model. As shown in Figure~\ref{fig:prior_pred}, the prior predictive distribution is diffuse: it ranges from nearly empty to nearly complete layers, so the prior does not constrain the statistics to a narrow region. Every observed statistic lies near the centre of its prior predictive distribution, between the $31$st and the $58$th percentiles, so the prior is compatible with the data without favouring them.

\subsubsection{Network structure}\label{sec:net_structure}

\begin{figure}[htbp]
   \centering
   \includegraphics[scale = 0.30]{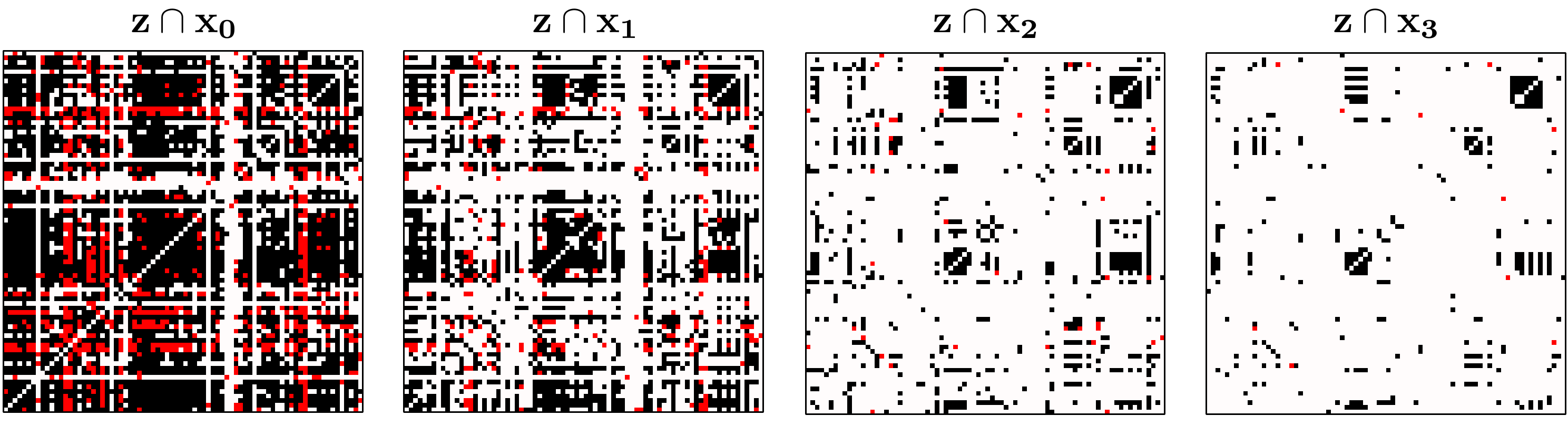}
   \caption{Signed adjacency matrices of the four layers $\bfx_0, \ldots, \bfx_3$ of the weighted adjacency matrix. Positive edges are shown as black pixels, and negative edges as red pixels.}
   \label{fig:adj_mats}
\end{figure}

The STRING network comprised $2{,}086$ interactions; after the exclusion of the two genes without co-expression data, the baseline layer contains $1{,}993$ edges among the 78 genes, a relatively dense graph (edge density $\approx 0.66$). Four adjacency matrices were constructed by applying STRING confidence score thresholds at $0.15$, $0.40$, $0.70$, and $0.90$, corresponding to the baseline layer $\mathbf{x}_0$ and the higher-confidence interaction layers $\mathbf{x}_1$ (medium confidence), $\mathbf{x}_2$ (high confidence), and $\mathbf{x}_3$ (highest confidence), respectively (see Figure~\ref{fig:adj_mats}). The number of edges decreased from $1{,}993$ in $\mathbf{x}_0$ to $1{,}100$ in $\mathbf{x}_1$, $379$ in $\mathbf{x}_2$, and $170$ in $\mathbf{x}_3$. The proportion of positive edges increases monotonically across layers, rising from $82.3\%$ ($1{,}640$ out of $1{,}993$) in $\mathbf{x}_0$ to $88.2\%$ ($970/1{,}100$) in $\mathbf{x}_1$, $94.7\%$ ($359/379$) in $\mathbf{x}_2$, and $97.1\%$ ($165/170$) in $\mathbf{x}_3$. At the baseline layer $\mathbf{x}_0$, a total of $32{,}092$ triangles were detected, of which $21{,}661$ ($67.5\%$) were of type \texttt{(+++)}, $5{,}933$ of type \texttt{(-++)}, $4{,}123$ of type \texttt{(+--)}, and $375$ of type \texttt{(---)}. As the confidence threshold increases, the triangle count drops sharply---to $9{,}239$, $1{,}081$, and $267$ in $\mathbf{x}_1$, $\mathbf{x}_2$, and $\mathbf{x}_3$, respectively---while the proportion of \texttt{(+++)} triangles grows substantially, reaching $80.6\%$, $93.3\%$, and $96.3\%$ at the three higher layers, suggesting that structural balance becomes increasingly dominant as lower-confidence interactions are removed.

\subsection{Results}\label{sec:app_results}

\subsubsection{Posterior inference}

Posterior inference was carried out with Algorithm~\ref{alg:ae} using the joint update. As anticipated in Section~\ref{sec:sim_design}, in the dense first two layers \texttt{gwesp+++} is close to $e^{\alpha_k}$ times \texttt{edges+}, so the posterior of $(\varphi_{k,\texttt{edges+}}, \varphi_{k,\texttt{gwesp+++}})$ is concentrated near a line: their posterior correlation is $-0.99$, $-0.96$ and $-0.77$ in the three layer transitions. The proposal covariance was matched to the shape of the posterior. It was initialised at $\mathbf{B}^{(0)} = \{\mathcal{I}(\hat{\bfvarphi}) + \bfLambda\}^{-1}$, where $\hat{\bfvarphi}$ is the maximum pseudo-likelihood estimate, $\mathcal{I}(\hat{\bfvarphi}) = \mathrm{Cov}_{\hat{\bfvarphi}}\{\bfs\}$ is the Fisher information estimated from $500$ networks simulated at $\hat{\bfvarphi}$, and $\bfLambda$ is the prior precision of $\bfvarphi$ implied by the hierarchical prior with $\bfSigma_{\bfvarphi}$ fixed at its prior mean, and it was then re-estimated from the chain during the burn-in.

Four chains were run in parallel for $40{,}000$ iterations each, one started at $\hat{\bfvarphi}$ and three at independent draws from $\mathcal{N}(\hat{\bfvarphi}, 4\mathbf{B}^{(0)})$. The initial scale was $\gamma = 2.38/\sqrt{2Kp} \approx 0.40$, the usual random-walk Metropolis scale \citep{rob:ros09} reduced by a factor $\sqrt{2}$ to allow for the additional noise of the exchange step. The scale and the proposal covariance were adapted every $500$ iterations between iterations $1{,}000$ and $8{,}000$, using the acceptance rate over the preceding $500$ iterations and $\lambda = 0.1$, and then held fixed; the first $8{,}000$ iterations were discarded as burn-in. Each auxiliary network was obtained by $20{,}000$ MCMC steps started at the observed network. Retaining every fourth iteration gave $8{,}000$ draws per chain ($32{,}000$ in total).

The acceptance rates after burn-in were between $0.22$ and $0.29$. The potential scale reduction factor \citep{gel:rub92} was at most $1.024$ for the layer-specific parameters and $1.003$ for the hierarchical means, and the effective sample sizes over the four chains were between $637$ and $1{,}128$ for the layer-specific parameters and between $2{,}241$ and $4{,}093$ for $\bfmu_{\bfvarphi}$. The largest posterior correlation between parameters of different layers was $0.15$, so for this model the layer-wise update with adapted covariance matrices $\mathbf{B}_k$ would also have been adequate. The posterior estimates, together with the potential scale reduction factor and the effective sample size of each parameter, are displayed in Table~\ref{tab:post_mu_phi}.

\begin{table}[h!]
\centering
\caption{Summary of MCMC posterior estimates of the global and layer-specific parameters; potential scale reduction factor ($\hat{R}$); Effective sample size (ESS) over the four chains ($32{,}000$ retained draws).}\label{tab:post_mu_phi}
\begin{tabular}{c|l|rrrr|rr}
\hline
Layer & Parameter & Mean & 2.5\% & 50\% & 97.5\% & $\hat{R}$ & ESS \\
\hline
\multirow{6}{*}{-}
 & $\mu_{\bfvarphi, \texttt{edges+}}$                & -1.373 & -2.072 & -1.380 & -0.629 & 1.001 & 2{,}290 \\
 & $\mu_{\bfvarphi, \texttt{nm(myo)}}$ & 0.028 & -0.393 & 0.028 & 0.441 & 1.000 & 4{,}093 \\
 & $\mu_{\bfvarphi, \texttt{nm(hall)}}$   & 0.171 & -0.258 & 0.167 & 0.618 & 1.000 & 3{,}030 \\
 & $\mu_{\bfvarphi, \texttt{gwesp+++}}$            & 0.672 & 0.199 & 0.671 & 1.161 & 1.001 & 3{,}060 \\
 & $\mu_{\bfvarphi, \texttt{gwesp+--}}$             & -0.174 & -0.705 & -0.165 & 0.305 & 1.001 & 2{,}241 \\
 & $\mu_{\bfvarphi, \texttt{gwesp-++}}$            & -0.969 & -1.408 & -0.970 & -0.528 & 1.003 & 2{,}437 \\
\hline
\multirow{6}{*}{1}
 & $\varphi_{1,  \texttt{edges+}}$                & -0.622 & -1.490 & -0.648 & 0.345 & 1.008 & 732 \\
 & $\varphi_{1,  \texttt{nm(myo)}}$ & -0.013 & -0.221 & -0.013 & 0.195 & 1.002 & 969 \\
 & $\varphi_{1,  \texttt{nm(hall)}}$   & 0.139 & -0.090 & 0.139 & 0.370 & 1.005 & 792 \\
 & $\varphi_{1,  \texttt{gwesp+++}}$            & 0.838 & 0.197 & 0.851 & 1.420 & 1.009 & 729 \\
 & $\varphi_{1,  \texttt{gwesp+--}}$             & 0.147 & 0.105 & 0.147 & 0.192 & 1.009 & 1{,}128 \\
 & $\varphi_{1,  \texttt{gwesp-++}}$            & -0.969 & -1.103 & -0.968 & -0.842 & 1.002 & 704 \\
\hline
\multirow{6}{*}{2}
 & $\varphi_{2,  \texttt{edges+}}$                & -2.310 & -3.079 & -2.291 & -1.652 & 1.004 & 656 \\
 & $\varphi_{2,  \texttt{nm(myo)}}$ & 0.119 & -0.171 & 0.122 & 0.416 & 1.005 & 736 \\
 & $\varphi_{2,  \texttt{nm(hall)}}$   & -0.069 & -0.371 & -0.068 & 0.220 & 1.001 & 713 \\
 & $\varphi_{2,  \texttt{gwesp+++}}$            & 0.979 & 0.674 & 0.967 & 1.354 & 1.003 & 637 \\
 & $\varphi_{2,  \texttt{gwesp+--}}$             & -0.392 & -0.834 & -0.374 & -0.060 & 1.007 & 882 \\
 & $\varphi_{2,  \texttt{gwesp-++}}$            & -1.207 & -1.557 & -1.202 & -0.881 & 1.024 & 817 \\
\hline
\multirow{6}{*}{3}
 & $\varphi_{3,  \texttt{edges+}}$                & -1.317 & -1.767 & -1.315 & -0.871 & 1.002 & 715 \\
 & $\varphi_{3,  \texttt{nm(myo)}}$ & -0.016 & -0.471 & -0.017 & 0.426 & 1.003 & 911 \\
 & $\varphi_{3,  \texttt{nm(hall)}}$   & 0.435 & -0.041 & 0.432 & 0.911 & 1.004 & 743 \\
 & $\varphi_{3,  \texttt{gwesp+++}}$            & 0.242 & 0.180 & 0.242 & 0.306 & 1.003 & 755 \\
 & $\varphi_{3,  \texttt{gwesp+--}}$             & -0.298 & -1.164 & -0.271 & 0.427 & 1.004 & 863 \\
 & $\varphi_{3,  \texttt{gwesp-++}}$            & -0.782 & -1.358 & -0.766 & -0.315 & 1.005 & 697 \\
\hline
\end{tabular}
\end{table}

\subsubsection{Interpretation of posterior estimates}

The posterior estimates indicate which structural features favour the persistence of positive and negative edges across interaction confidence layers.

The \texttt{gwesp+++} parameter is positive and credible in all three transitions and at the hierarchical level. Positive edges whose endpoints share positive partners are therefore more likely to persist at a higher confidence level, so that the cohesive positive core of the network is preferentially preserved as the threshold increases. The smaller value in the third transition is not directly comparable with the first two, because the statistic is on a different scale: with $\alpha_3 = 2$ an edge can contribute up to $e^2 \approx 7.4$, against $1.5$ and $1.8$ in the first two transitions. For the same reason, the hierarchical means of the geometrically weighted terms average coefficients that multiply statistics on different scales, and should be read as summaries of sign and approximate magnitude rather than as a common effect size; the layer-specific estimates are the primary quantities of interest for these terms.

The \texttt{edges+} parameter refers to positive edges with no positive shared partners, for which the \texttt{gwesp+++} contribution is zero, and is negative in all three transitions, credibly so in the second and third: positive edges that are not embedded in any positive triangle tend to dissolve. In the first two transitions, \texttt{edges+} and \texttt{gwesp+++} are separately identified only through the relatively few positive edges with no or few positive shared partners, which explains the wide intervals of $\varphi_{1,\texttt{edges+}}$ and $\varphi_{1,\texttt{gwesp+++}}$ and their strong negative posterior correlation. The log-odds contribution for a positive edge with many positive shared partners is much more precisely estimated: $0.628$ ($[0.474, 0.787]$) in the first transition and $-0.525$ ($[-0.732, -0.317]$) in the second.

The homophily effects based on Myogenesis membership and functional pleiotropy are not credibly different from zero in any transition or at the hierarchical level. The pleiotropy effect in the third transition has the largest posterior mean among these parameters ($0.435$), but its credible interval also includes zero ($[-0.041, 0.911]$). The data therefore provide no evidence that either covariate affects the persistence of positive edges in this network.

The \texttt{gwesp+--} parameter is positive and credible in the first layer transition, from baseline to medium confidence ($0.147$, $[0.105, 0.192]$), negative in the second ($-0.392$, $[-0.834, -0.060]$) and uncertain in the third. In the first transition, positive edges whose endpoints share negative partners tend to persist, consistent with local structural balance; in the second they tend to dissolve, and the configuration is almost absent from $\bfx_2$ (the statistic falls from $444$ in $\bfx_1$ to $13$ in $\bfx_2$).

The \texttt{gwesp-++} parameter is negative and credible in all three transitions ($-0.969$, $-1.207$ and $-0.782$) and at the hierarchical level ($-0.969$, $[-1.408, -0.528]$). Negative edges whose endpoints share positive partners, which form unbalanced \texttt{(-++)} triangles, are therefore preferentially dissolved as the confidence threshold increases, in agreement with structural balance theory and with the progressive disappearance of unbalanced triangles across layers reported in Section~\ref{sec:net_structure}. The effect remains strongly negative when \texttt{gwesp+++} is included in the model, so it is not a by-product of closure among positive edges. 

From a biological perspective, high-confidence interactions between anti-correlated genes are thus rarely embedded in a positively co-expressed neighbourhood, consistent with anti-correlated genes operating in distinct functional modules rather than as components of the same regulatory circuit \citep{barabasi:2004}. Since the effect persists in the highest-confidence transition, it is unlikely to be an artefact of low-confidence or indirect interaction evidence. The mechanism underlying the effect cannot be identified from the present data; a possible explanation is topological, as the 78 genes are effectors of contiguous cardiac functions and most of their interactions are positive ($82.3\%$ of the edges of $\mathbf{x}_0$), so that negative edges embedded in positive neighbourhoods are atypical of the module.

The posterior standard deviations of the layer-specific effects around the hierarchical means have posterior means between $0.32$ and $0.50$, close to the prior expectation of about $0.35$. As shown in Section~\ref{sec:sim_results}, with $K = 3$ layers, the between-layer covariance is largely determined by the prior, and the layer-specific estimates are partially pooled towards the hierarchical means.

\subsubsection{Posterior predictive diagnostics}

\begin{figure}[h]
   \centering
   \includegraphics[width = \textwidth]{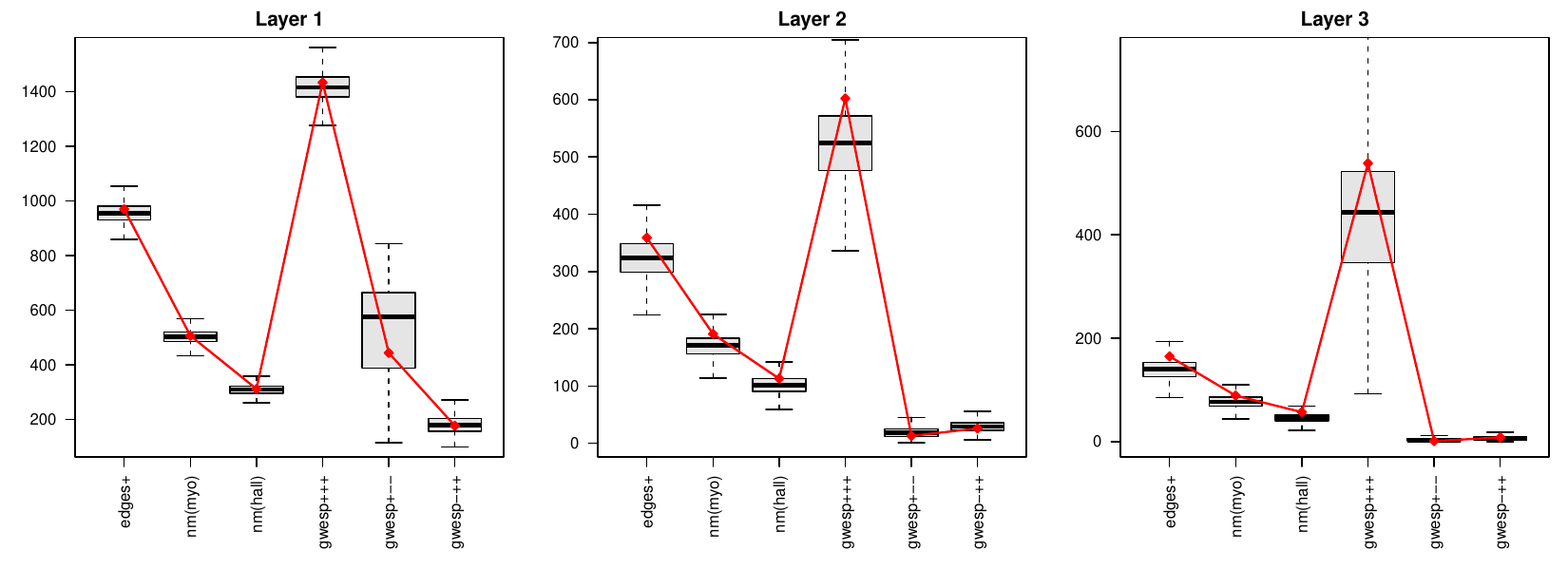}
   \caption{Posterior predictive distributions of the six statistics of each layer transition, based on $1{,}000$ simulated networks and summarised by boxplots (outlying values not shown), with the observed values indicated by red diamonds connected by a solid red line.}
   \label{fig:post_pred}
\end{figure}

Posterior predictive diagnostics assess the adequacy of the multi-layer dissolution ERGM in reproducing the observed signed weighted network. We simulated $1{,}000$ parameter vectors from the posterior sample and simulated one network for each, as for the prior predictive checks. As shown in Figure~\ref{fig:post_pred}, all the network statistics lie within the central $95\%$ posterior predictive intervals, and the proportions of simulated values below the observed ones range from $0.12$ to $0.88$, so none of the statistics is systematically missed. The largest deviations are a mild under-prediction of the number of positive edges in the second and third transitions, which may reflect the partial pooling of the layer-specific parameters towards the hierarchical means described above.

\section{Conclusions}\label{conc}

In this paper, we introduce a novel modelling framework for signed weighted networks based on the multi-layer dissolution ERGM processes and adopt the model separability of conditional signed weighted processes given the interaction structure. Our model jointly captures the dynamics of dyadic interactions across layers, defined according to different levels of edge strength and the persistent sign of relationships. By linking layer-specific dyads to the overall sign of the relationship through cumulative change statistics, the framework quantifies how positive or negative interactions reinforce strong edges between nodes.

This approach is highly interpretable: changes in network statistics at the layer level translate directly into the probability of observing a positive, negative, or inactive edge in higher layers, while the aggregation of information across active layers informs the overall sign of the relationship.
This framework enables researchers to uncover underlying structural mechanisms, such as balance and cohesion, and to distinguish transient interactions from enduring relational tendencies. Importantly, by incorporating both layer-specific and joint parameters, the model accommodates heterogeneity in dyadic behaviour across contexts.

Our fully probabilistic implementation, via Bayesian hierarchical inference, allows for the estimation of both global effects and the extent of heterogeneity across layers. Posterior inference is implemented using an adaptive exchange algorithm that updates the layer-specific parameters either layer by layer or jointly, with proposal covariances estimated during the burn-in. Learning the proposal covariance proved essential in the application: in dense layers, geometrically weighted closure statistics with small decay are nearly proportional to the edge count and an isotropic proposal explores it very slowly. In such layers, the closure and edge parameters are identified separately only through the edges with few shared partners, while their combined effect on well-embedded edges is precisely estimated, and results should be reported accordingly.

A simulation study showed that the adaptive exchange algorithm recovers the layer-specific parameters with little bias and well-calibrated credible intervals. The remaining bias is consistent with the shrinkage towards the common mean induced by the hierarchical prior. The study also showed that signed triadic effects are identifiable only when the network contains enough triangles of the corresponding type, which can be checked before fitting through the Fisher information of the model at plausible parameter values. With only three layers, the prior largely determines the between-layer covariance, so layer-specific contrasts should be read with this in mind.

The model is implemented in R \citep{R}, based on the \textbf{ergm} \citep{ergm4} and \textbf{ergm.multi} \citep{multi.ergm} packages. The \textbf{Bergm.ws} package which was used to obtain the results presented in this study, is publicly available at \url{https://github.com/acaimo/Bergm.ws}.

Looking ahead, the proposed approach provides a principled basis for modelling temporally evolving, signed weighted networks, in which both the strength and the polarity of interactions may vary over time. In fact, by explicitly integrating temporal dynamics, for instance, through separable temporal models \citep{kri:han14,kei:etal23,cai:gol26}, multi-layer dissolution ERGMs can disentangle the distinct mechanisms that strengthen, persist, or dissolve positive and negative edges over time, while allowing  rigorous inference and prediction of how overall complex relational structures evolve.

\subsection*{Acknowledgements}
The authors sincerely thank Prof. Adamo Pio d'Adamo for supporting us in our biological and genetic interpretation throughout this work.

\subsection*{Data and code availability}
All data analysed in this study were derived from publicly available resources: Jensen Diseases database, STRING (version 12.0), ARCHS4 human gene correlation matrix (version 1.0) and MSigDB Hallmark collection. The \textbf{Bergm.ws} package which was used to obtain the results presented in this study, is publicly available at \url{https://github.com/acaimo/Bergm.ws}

\subsection*{Declaration of AI use in the manuscript preparation process}
During the preparation of this work the authors used Claude (Anthropic) in order to check the internal consistency of the manuscript and to revise the wording of selected passages. After using this tool, the authors reviewed and edited the content as needed and take full responsibility for the content of the publication.

\subsection*{Funding}
This publication has emanated from research conducted with the financial support of Taighde Éireann - Research Ireland, under Grant number [23/RC/13506] at Rinn Artificial Intelligence.

\bibliographystyle{abbrvnat}

\label{lastpage}

\end{document}